\documentclass[journal]{IEEEtran}
\IEEEoverridecommandlockouts

\usepackage{cite}
 \usepackage[pdftex]{graphicx}
 
\DeclareGraphicsExtensions{.pdf,.jpeg,.png}
\graphicspath{.pdf/}

\begin{document}

\title{TRITIUM - A Real-Time Tritium Monitor System for Water Quality Surveillance}

\author{C.D.R.~Azevedo, A.~Baeza, M.~Br\'as, T.~C\'amara, C.~Cerna, E.~Chauveau, J.M.~Gil, J.A.~Corbacho,V.~Delgado, J.~D\'iaz, J.~Domange, C.~Marquet, M.~Mart\'inez-Roig, A.~Moreno, F.~Piquemal, A.~Rodr\'iguez, J.~Rodr\'iguez, C.~Rold\'an, J.F.C.A.~Veloso, N.~Yahlali 
\thanks{T.~C\'amara, V. Delgado, J.~D\'iaz, M.~Mart\'inez-Roig and N. Yahlali (corresponding author) are with Instituto de F\'isica Corpuscular, Centro mixto CSIC-Universidad de Valencia, Paterna, Spain. (email: nadia.yahlali@ific.uv.es)}
\thanks{C.D.R.~Azevedo, M.~Br\'as and J.F.C.A.~Veloso are with I3N - Departamento de Fisica da Universidade de Aveiro, Aveiro, Portugal} 
\thanks{C.~Rold\'an is with Universidad de Valencia, Departamento de F\'isica Aplicada, Burjassot, Spain.}
\thanks{A.~Baeza, J.A.~Corbacho and A.~Rodr\'iguez are with Universidad de Extremadura, Laboratorio de Radioactividad Ambiental. Servicio de apoyo a la     investigaci\'on, C\'aceres, Spain}
\thanks{C.~Cerna, E.~Chauveau, J.~Domange, C.~Marquet and F.~Piquemal are with Universit\'e de Bordeaux and CNRS, CENBG, Gradignan Cedex, France.} 
\thanks{J.M.~Gil, A.~Moreno and J.~Rodr\'iguez are with Junta de Extremadura, Plaza del Rastro s/n. 06800, M\'erida, Spain.}
}
\maketitle

\IEEEpubid{978-1-5386-8495-5/18\$31.00~\copyright~2018 IEEE}

\begin{abstract}
In this work the development results of the TRITIUM project is presented. The main objective of the project is the construction of a near real-time monitor for low activity tritium in water, aimed at in-situ surveillance and radiological protection of river water in the vicinity of nuclear power plants. The European Council Directive 2013/51/Euratom requires that the maximum level of tritium in water for human consumption to be lower than 100 Bq/L. 
Tritium levels in the cooling water of nuclear power plants in normal operation are much higher than the levels caused by the natural and cosmogenic components, and may easily surmount the limit required by the Directive. 
The current liquid-scintillation measuring systems in environmental radioactivity laboratories are sensitive to such low levels, but they are not suitable for real-time monitoring.  Moreover, there is no currently available device with enough sensitivity and monitoring capabilities that could be used for surveillance of the cooling water of nuclear power plants.  A detector system based on scintillation fibers read out by photomultiplier tubes (PMTs) or silicon photomultiplier (SiPM) arrays is under development for in-water tritium measurement. This detector will be installed in the vicinity of Almaraz nuclear power plant (Spain) in Spring 2019. An overview of the project development and the results of first prototypes are presented. 
\end{abstract}
\IEEEpubidadjcol

\IEEEpeerreviewmaketitle

\section{Introduction}

\subsection{The SUDOE TRITIUM Project}

The TRITIUM project is funded by the INTERREG-SUDOE program of the European Union and composed by a consortium of 5 southwestern european institutions: University of Extremadura (Spain), University of Valencia (Spain), University of Aveiro (Portugal), CENBG Bordeaux (France) and Junta de Extremadura (Spain). The main objective of the project is to construct a monitor for low activity tritium with near real-time capability, for in-situ surveillance and radiological protection of river water in the vicinity of nuclear power plants. The EU Council Directive 2013/51/Euratom requires the maximum level of tritium in water for human consumption to be lower than 100 Bq/L. These levels are much higher than those caused by the natural or cosmogenic components, however they may easily surpass the limit required by the Directive in the cooling water of nuclear power plants in normal operation. This water is usually released in rivers after cooling and may cause direct or indirect contamination of the environment and water for human consumption with tritium, a long-lived hydrogen isotope. Monitoring the tritium level in the cooling water of nuclear power plants would provide a safety alarm and an indicator of anomalous functioning due to an excess of neutrons or a water leakage from the primary cooling circuit.  Indeed, tritium is the third isotope of hydrogen produced by neutron capture of hydrogen or deuterium atoms, thus an excess of neutrons would induce an excess of tritium level in the cooling water. In order to measure low levels of tritium with the sensitivity required by the european Directive, the consortium is developing a detector system based on scintillating optical fibers, with active and passive cosmic background rejection. The installation of the detector is planned in Spring 2019 in the Tagus river, downstream the dam of the Almaraz (Spain) nuclear power plant, near the Portuguese border.
\IEEEpubidadjcol

\subsection{State of the Art}
Tritium level measurement is one of the systematic controls to be carried out during energy production by nuclear fission and, in the future, by nuclear fusion. Tritium is an hydrogen isotope with 12 years half-life, decaying into $^3$He through beta emission. The emitted electron has a continuous energy spectrum peaked at 3 keV with a maximum energy emission of 18 keV, resulting in an averaged energy of 6 keV. Due to such low energy, the detection of tritium decay is not trivial and requires high sensitivity detectors. There are already available methods for the tritium detection: gaseous detectors, semi-conductors and liquid/solid scintillation. The gaseous detection method consists on the introduction of water vapor into an ionisation chamber, resulting in a too high complex system for autonomous operation. Solid state detectors, such APDs, can be directly exposed to the beta radiation but the presence of water makes them unusable. Liquid scintillators have been extensively used in laboratory to detect tritium levels in water as low as 1 Bq/L, but they require collection of water samples and data taking during several days, which makes them not suitable for real-time monitoring. Furthermore, they produce highly toxic residues (as toluene), forbidding their use for environmental surveillance. Solid scintillators have been suggested by different authors: K.J. Hofstetter in 1993~\cite{Hofstetter} built a capillary system read out by PMTs obtaining a detection limit value of 22000 Bq/L in 2 min of measurement. Another suggestion can be pointed to Rathnakaran in 2000~\cite{Rathnakaran} with detection limits of 10000 Bq/L during 10 min of measurement, a number with 2 orders of magnitude higher than the required for this project. 
An alternative to solid scintillators, which provides higher sensitivity to low levels of tritium in water, is the use of scintillating optical fibers as a detection medium. This scintillation alternative, which is proposed in the present work, was investigated in 1998 by J.W. Berthold and L.A. Jeffers \cite{Berthold}, using polystyrene optical fibers with fluor-doped cladding read out by photomultiplier tubes (PMTs). These authors aimed at the development of a detector for monitoring tritium in drinking waters to verify compliance with the US EPA drinking water standard of 740 Bq/L, a much higher limit than that required by the European Directive.  

\section{TRITIUM System Technology}

The design of TRITIUM detector is based on scintillating fibers with fluor-doped core, without cladding, in order to maximize the detection area exposed to water, due to the small energy (6~keV on average) of electrons from tritium decay to be measured. Scintillating fibers of 1~mm and 2~mm diameter have been considered in the R\&D phase, while for light detection the options in study are PMTs or silicon photomultipliers (SiPMs). These latter have the advantage of low polarisation voltages ($\sim$50 V instead of typically $\sim$1 kV for PMTs), robustness and easy temperature dependance correction that may be implemented automatically. On the other hand, using PMTs has the advantage of a lower number of electronic channels and a larger detection area. In both options, the photodetectors reading out a fiber-bundle will work in coincidence mode in order to reject their intrinsic noise (or self-emission in the case of PMTs), and in anti-coincidence mode with a cosmic discrimination system based on plastic scintillators. The water supplied to the detector, has to undergo a thorough purification process in order to avoid maintenance of the detector due to organic and mineral depositions on the fibers and the water-vessel inner surfaces. The water purification system provides levels of purity specified as hyperpure with a conductivity of 10~$\mu$S/cm. Furthermore, it has to preserve the in-water tritium level prior to the purification process, and ensure an adequate water flow in the detector, allowing the renewal of its water content every 10 min. The detector system has to provide an alarm signal in case of tritium level in water above 100 Bq/L, after autonomous and continuous measurements of less than 10 min.  

\section{Detector Development}
  
\subsection{Simulations}

\begin{figure}[!t]
\centering
\includegraphics[width=3.5in]{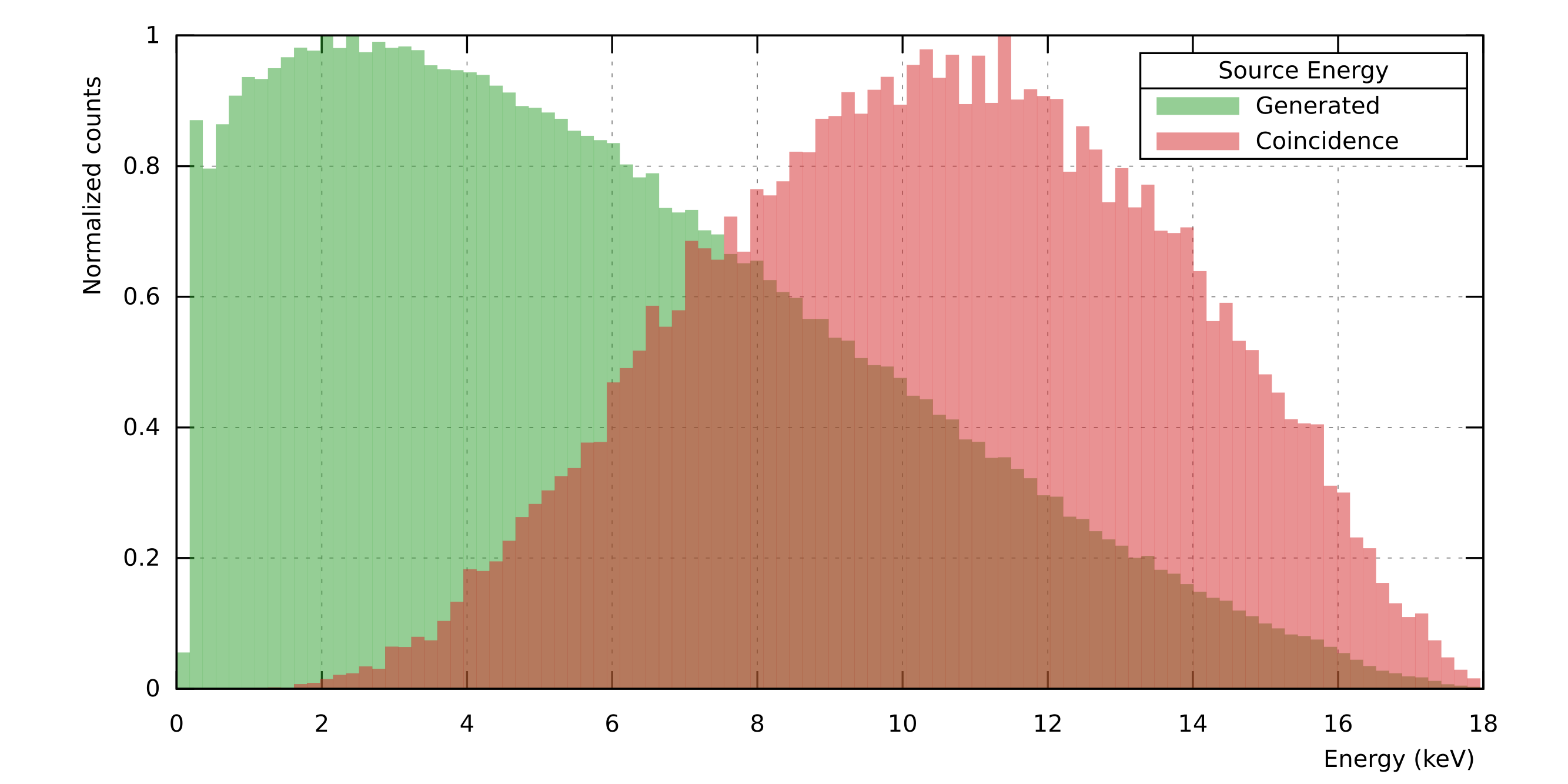}
\caption{Energy distribution of electrons from tritium decay (green) and their energy deposited in a scintillation fiber of 2~mm diameter (red) immersed in tritiated water.}
\label{figure-1}
\end{figure}

\begin{figure}[!t]
\centering
\includegraphics[width=3.5in]{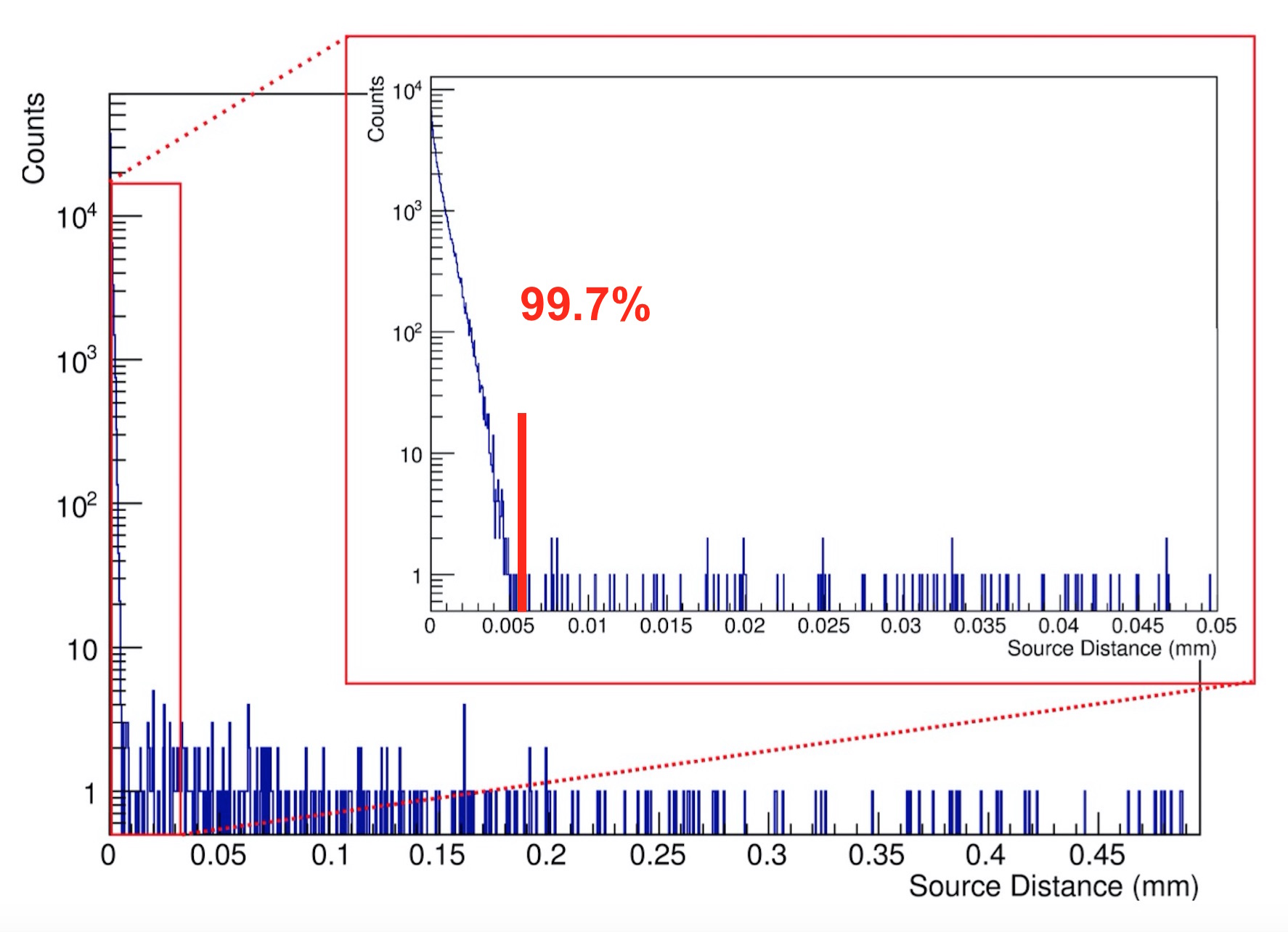}
\caption{Distribution of beta-electron hits in a scintillation fiber immersed in tritiated water as a function of the distance between the beta source and the fiber surface. 99.7\% of the events reaching the fiber surface are produced at a distance less then 5~$\mu$m from it. }
\label{figure-2}
\end{figure}

The proposed detector design using scintillating fibers of 1~mm or 2~mm diameter has been extensively studied through GEANT4 simulations~\cite{Geant4}.
Due to the low energy of the electrons emitted by tritium, and considering that the measurement medium is water, one can foresee that just a small part of the decays will be detected, as the majority of electrons will loose their energy in water. 
The relationship between the energy deposited by electrons in the fiber and the distance from their emission point to the fiber surface was studied by simulating a 2~mm fiber inside a water cylinder (of 0.5 mm thickness)
containing tritium uniformly distributed in the water volume.
The electron emission was considered to be isotropic and the energy was sampled from the tritium emission spectrum (Fig.~\ref{figure-1}, green distribution). The results are shown in Fig.~\ref{figure-2}, where it is observed that the majority (99.7\%) of events that reach the fiber surface are emitted from a distance less than 5~$\mu$m from it. This means that we can only detect the decays occurring in a very thin layer of water in the vicinity of the fiber surface, and that the 
scintillating optical fibers cannot have any cladding, as this usually exceeds by far 5~$\mu$m thickness in commercial scintillating fibers. Furthermore, the water to be measured in the detector should be free from organic or mineral residues that may deposit on the surface of the fibers. This requirement made necessary the development of a water purification system as part of the TRITIUM detection system, which ensures a stable operation of the detector and its long-term durability. 
\\
The energy distribution of electrons that reach the fiber surface was also studied, with the results presented in Fig.~\ref{figure-1}, where, in green, is the tritium emission spectrum used in these calculations, and in red, the electron energy distribution in the fiber, normalized to the emission spectrum in order to be easily compared. 
The shift of the energy peak of the electrons in the fiber, indicates, that the less energetic electrons will not reach the fibers, as they loose their energy in the water. 
It must be pointed out that the spectrum of energy deposition in the fiber shows the electrons that reach the fiber surface, which does not imply that they will produce enough photons to be detected.

\begin{figure}[!t]
\centering
\includegraphics[width=3.5in]{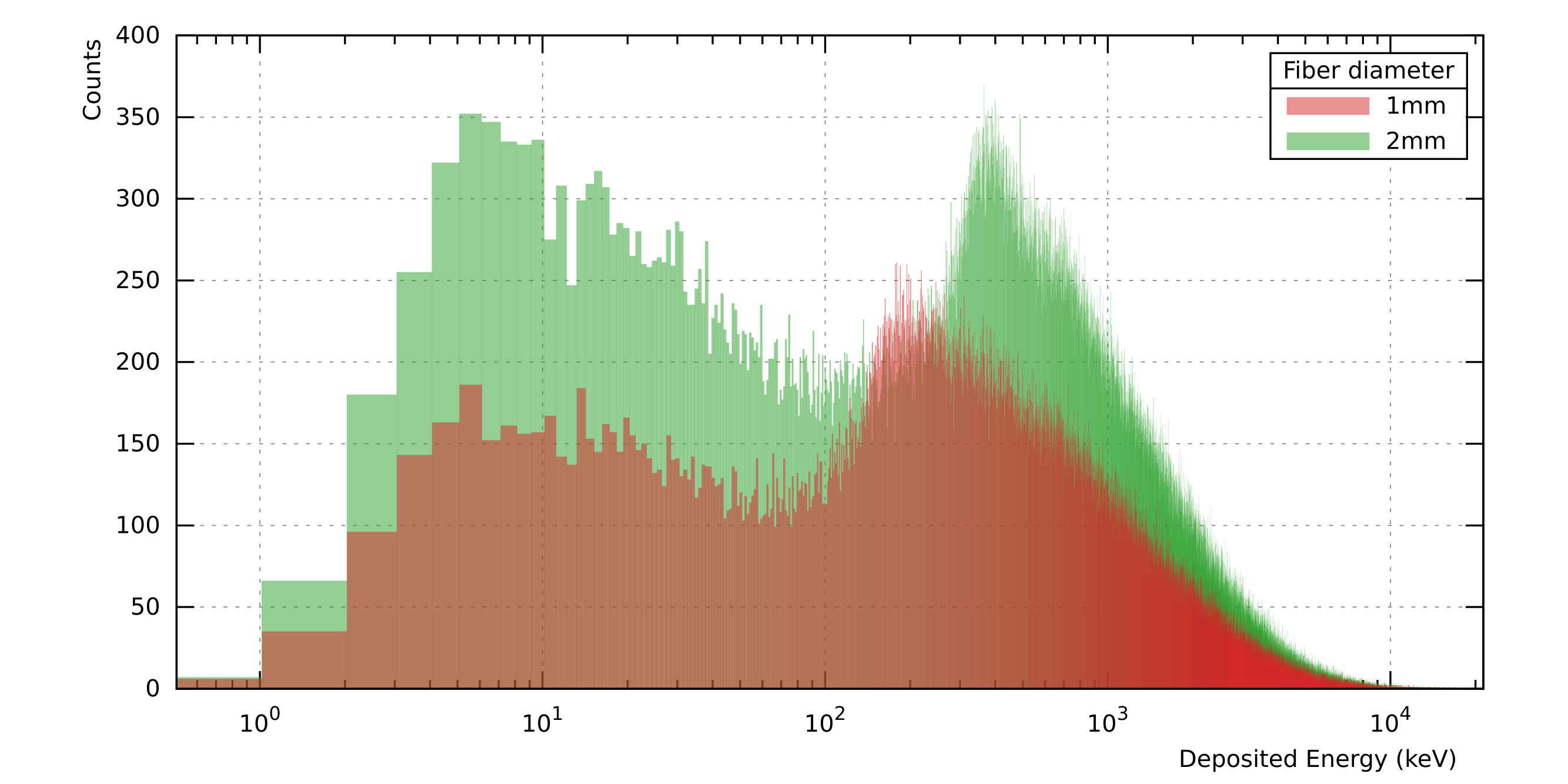}
\caption{Distribution of cosmic muon events as a function of the deposited energy in fibers of one millimeter (red) and two millimiter (green) diameter.}
\label{figure-3}
\end{figure}

Another important issue of the detector design is the diameter of the scintillating optical fibers. These are determined by a compromise between the detection efficiency of the fibers related to their light yield and collection per unit of energy deposited, and the amount of energy deposited from the background (cosmic muons essentially), which should be as low as possible. In Fig.~\ref{figure-3}, the energy deposited by cosmic muons in fibers of 1~mm and 2~mm diameter (red and green spectra respectively) are compared. As expected, the counting rate of cosmic muons in a fiber of 1~mm diameter is smaller than in 2 mm diameter in the whole energy range, and specially in the region of interest of tritium ($< 10$~keV). Therefore, scintillating fibers of 1 mm diameter are favoured due to the necessity of background rejection for increasing the detector sensitivity to low tritium activity. Furthermore, it is essential to reduce strongly the cosmic and natural radioactivity backgrounds in the detector, with cosmic VETO detectors. This active shield will be made out of scintillator detectors read out by photosensors. The whole detection system will be enclosed inside a lead castle of several cm thickness that will, in addition, reduce cosmic muons and natural radioactivity. 

\subsection{Prototypes}

\begin{figure}[!t]
\centering
\includegraphics[width=2.8in]{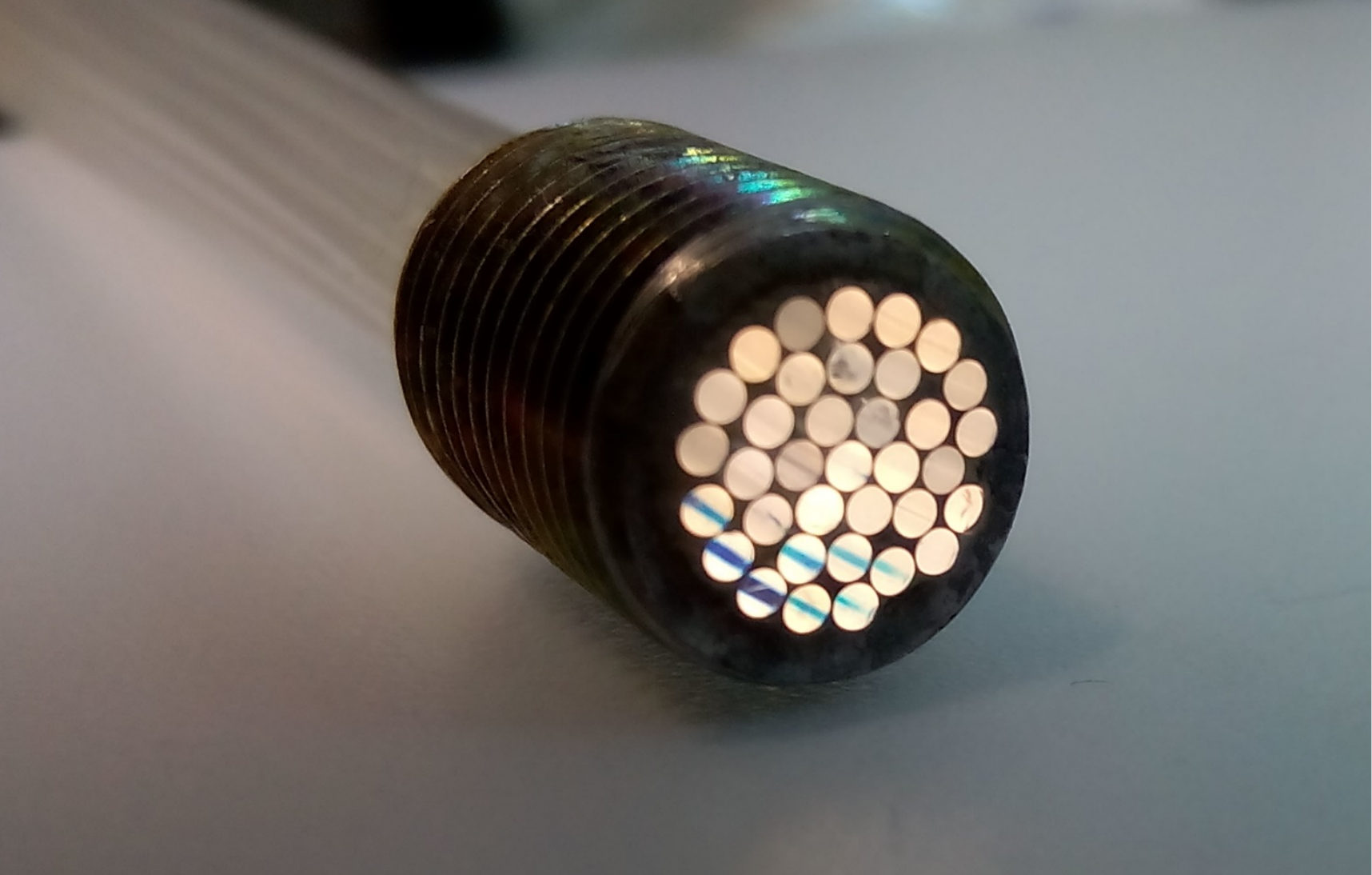}
\caption{One of the two end-connectors of the fiber-bundle of TRITIUM-0 prototype. The fiber ends are polished and coupled to the PMT with optical grease BC600 from Saint-Gobain Crystals~\cite{Saint-Gobain}. }
\label{figure-4}
\end{figure}

\begin{figure}[!t]
\centering
\includegraphics[width=2.8in]{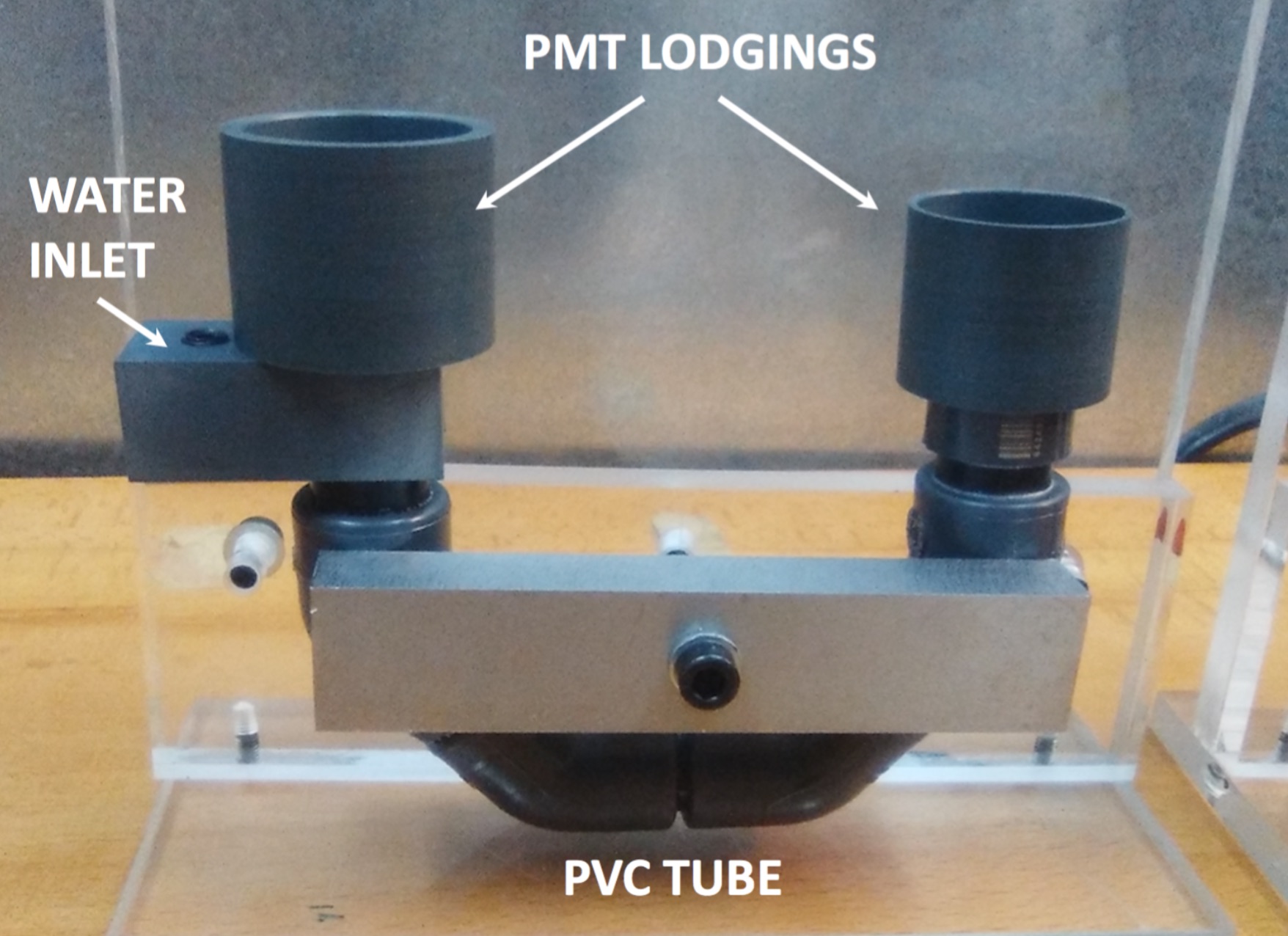}
\caption{Proof-of-principle prototype TRITIUM-0 instrumented with two PMTs. The fiber-bundle in the PVC tube was composed of 35 scintillating fibers BCF-12 of 1~mm diameter, 20 cm long and without cladding, from Saint-Gobain Crystals.}
\label{figure-5}
\end{figure}

Several TRITIUM prototypes were built for the detector proof of concept and sensitivity studies. The first prototype, called TRITIUM-0, was build using 35 scintillating optical fibers of 1 mm diameter, without cladding, from Saint-Gobain Crystals~\cite{Saint-Gobain}, which were terminated by the connector shown in 
Fig.~\ref{figure-4}. The fiber-bundle was introduced into a PVC tube with two bendings that allowed vertical coupling of the fibers to two PMTs, as depicted in Fig.~\ref{figure-5}. Optical grease BC630 from Saint-Gobain Crystals was used to optimise the light transmission to the photosensors. This prototype was first filled with 40 ml of hyperpure water for background measurements, and afterwards, it was filled with hyperpure highly tritiated water (108.11 MBq/L) for tritium activity measurement. This prototype that proved feasibility of the detection principle with scintillating fibers, put in evidence the important issues related to the light production and collection in the detector. 


\begin{figure}[!t]
\centering
\includegraphics[width=3.4in]{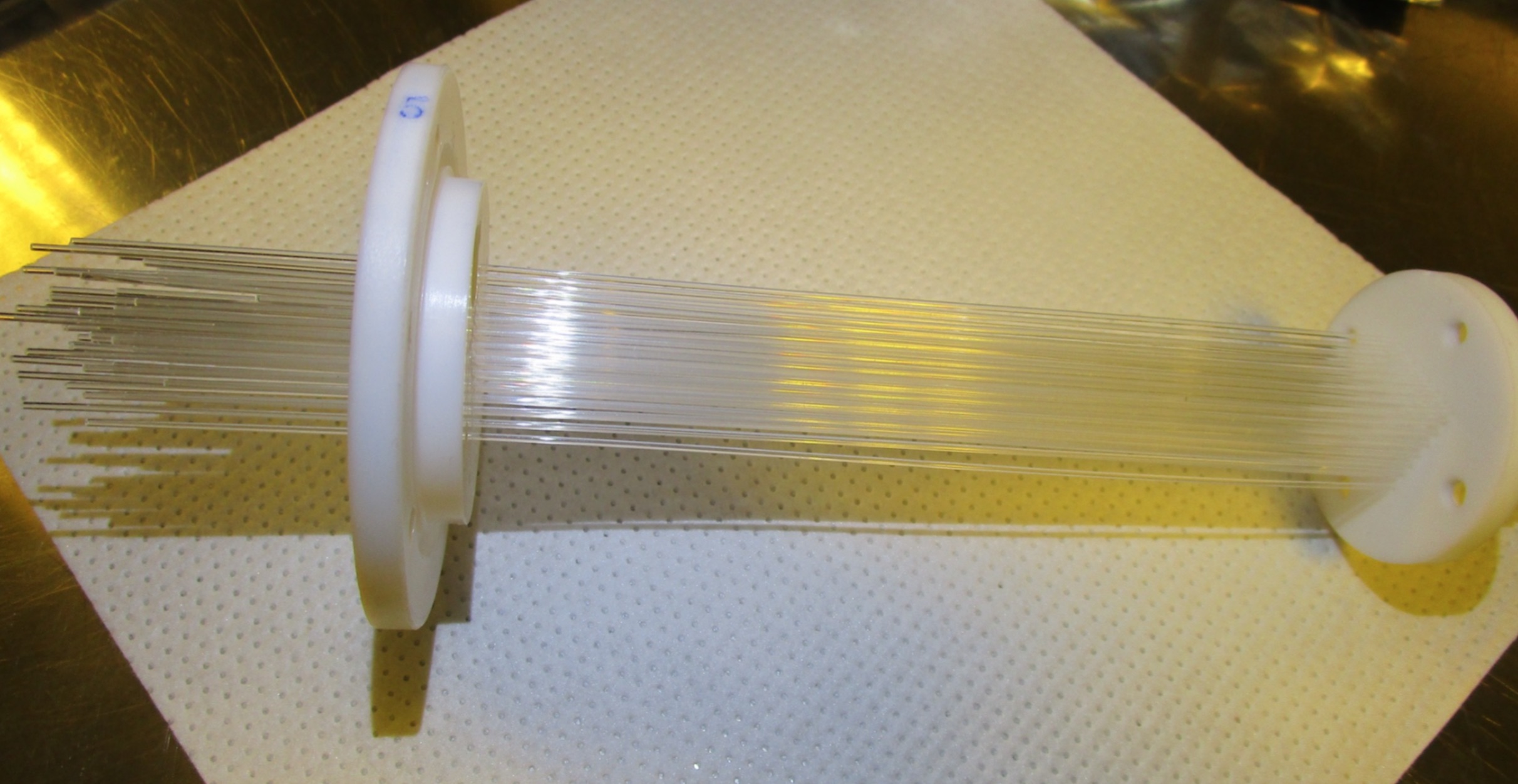}
\caption{Scintillating fibers mounted into two PTFE matrices used as top and bottom covers of TRITIUM-1 water vessel. The fibers end were polished and their surface conditioned in a clean room.}
\label{figure-6}
\end{figure}

\begin{figure}[!t]
\centering
\includegraphics[width=3.4in]{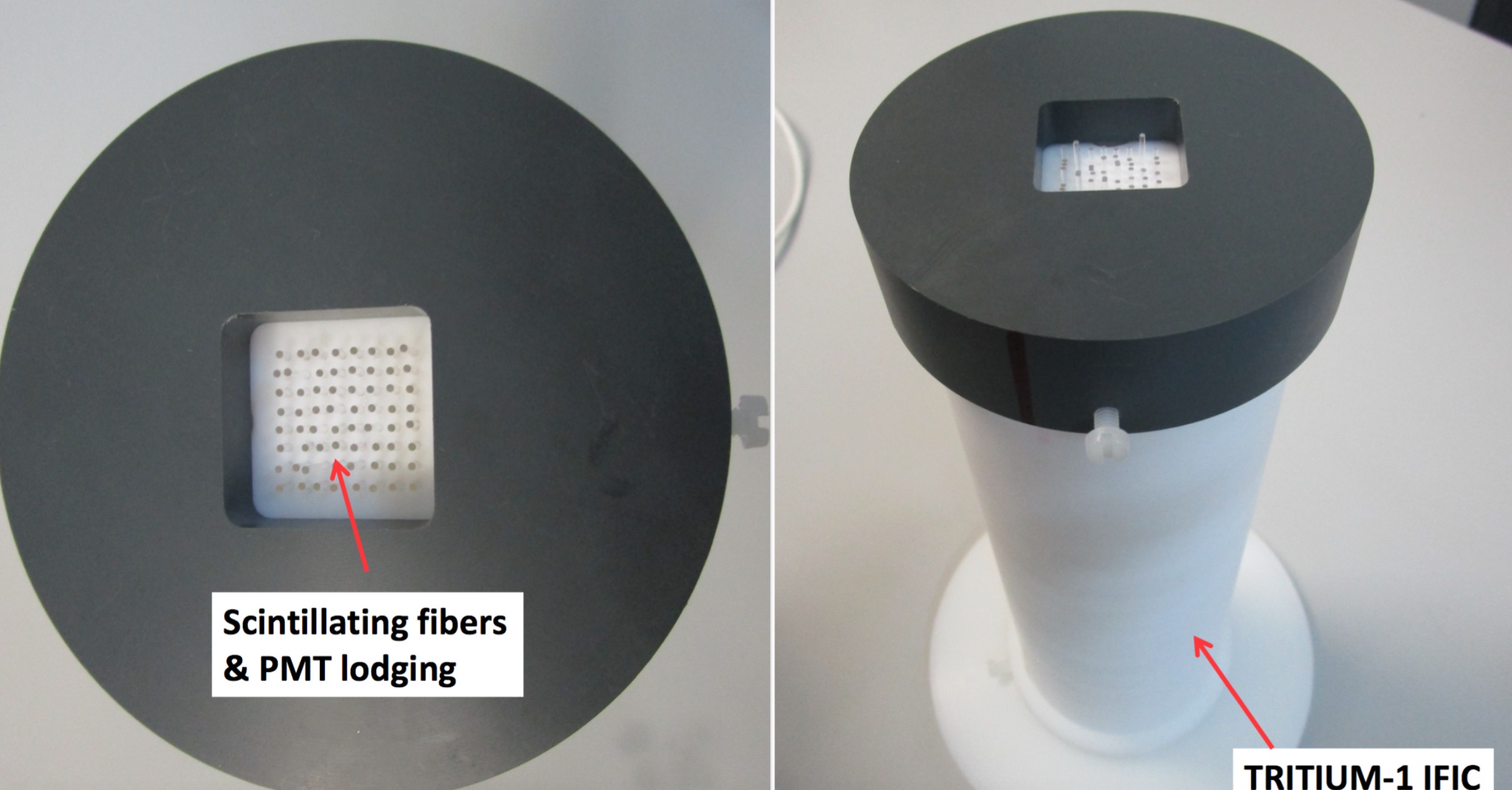}
\caption{TRITIUM-1 IFIC prototype with 64 scintillating fibers lodged vertically in a PTFE vessel (right picture), on top of which a PMT was placed in contact with the fibers. The prototype was placed in a dark chamber for data taking.}
\label{figure-7}
\end{figure}

Two other prototypes, called TRITIUM-1, were built at IFIC and Aveiro for light collection studies, long-term stability studies of scintillation fibers in water, readout and mechanical studies, including implementation of water flow solutions inside the vessel, water-tightness and optical coupling solutions for the photosensors (PMTs or SiPM arrays). IFIC prototype was composed of 64 scintillating fibers BCF-12 from Saint-Gobain Crystals, of 1 mm diameter without cladding, lodged in a highly reflective PTFE vessel. A protocol for conditioning of fibers was developed to optimize their in-water detection and light collection capability. This includes cleaving, polishing, and surface conditioning in a clean room, which ensures optimal cleanliness and wetting conditions (optimal contact with water) of the fibers surface. The fibers were then introduced into two PTFE matrices, as shown in Fig.~\ref{figure-6},  that allowed a separation of 1~mm between them and maintaining them in vertical position in the PTFE vessel shown in Fig.~\ref{figure-7}. On top of the vessel, a Hamamatsu R8520-406 PMT was lodged and put into contact with the fibers using optical grease. The PMT signal was preamplified, amplified and acquired in an AmpteK MCA-8000D digital Multichannel Analyzer.  

\begin{figure}[!t]
\centering
\includegraphics[width=3.5in]{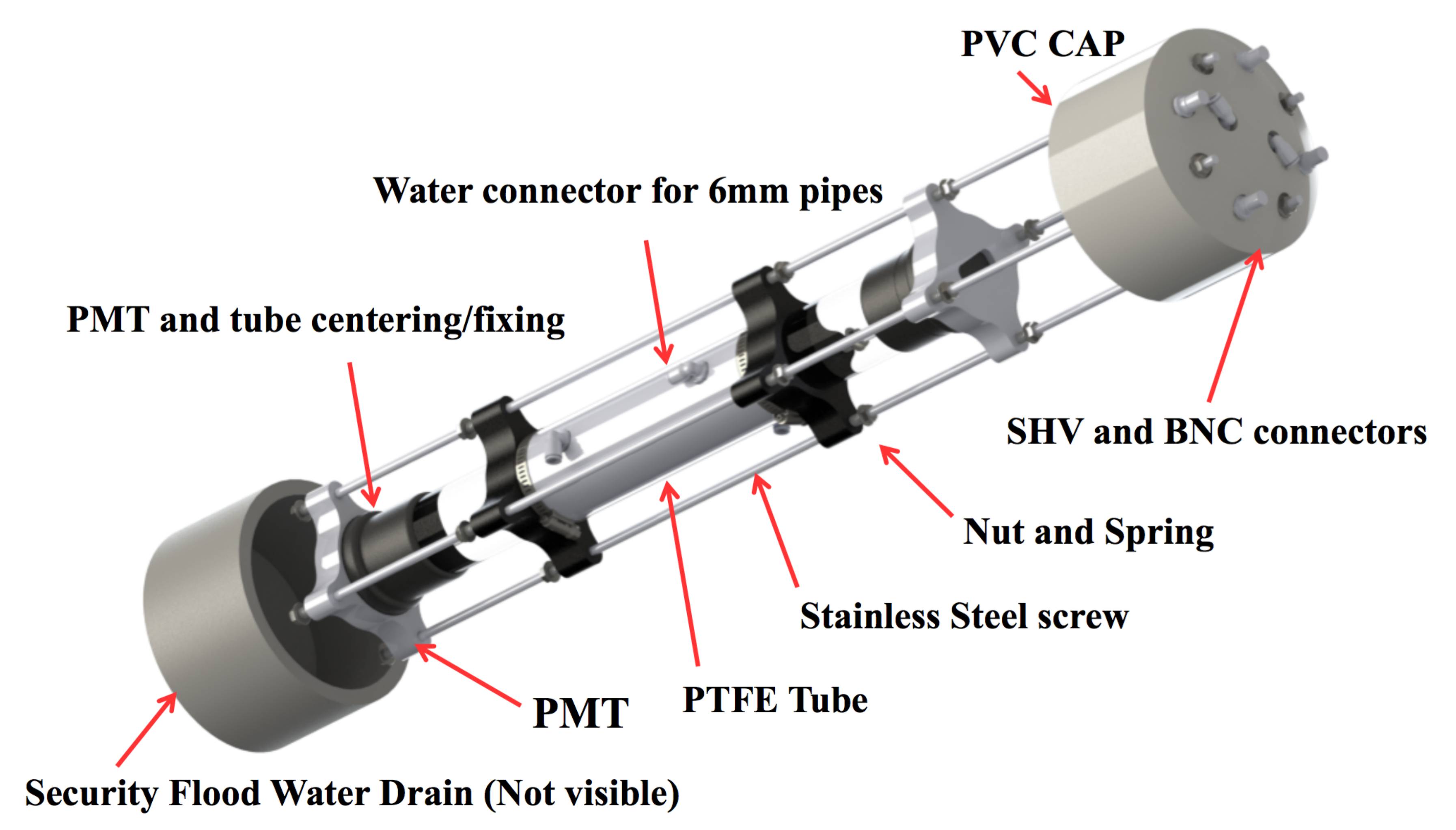}
\caption{TRITIUM-1 Aveiro prototype with 400 scintillating fibers, read out by two PMTs in coincidence mode. The fibers are lodged in a PTFE vessel closed by two PMMA windows which transmit the scintillation light to the PMTs. An inlet and an outlet for water flow are foreseen in the vessel body.}
\label{figure-8}
\end{figure}

A second version of TRITIUM-1, built in Aveiro, used also a PTFE vessel within which a high number (400) of tightly packed scintillating fibers was introduced. These were read out in coincidence mode by two Hamamatsu R2154-02 PMTs, through transparent PMMA windows of the water-vessel. A drawing of this prototype is shown in Fig.~\ref{figure-8}. 

\section{Preliminary Experimental Results}

\begin{figure}[!t]
\centering
\includegraphics[width=3.5in]{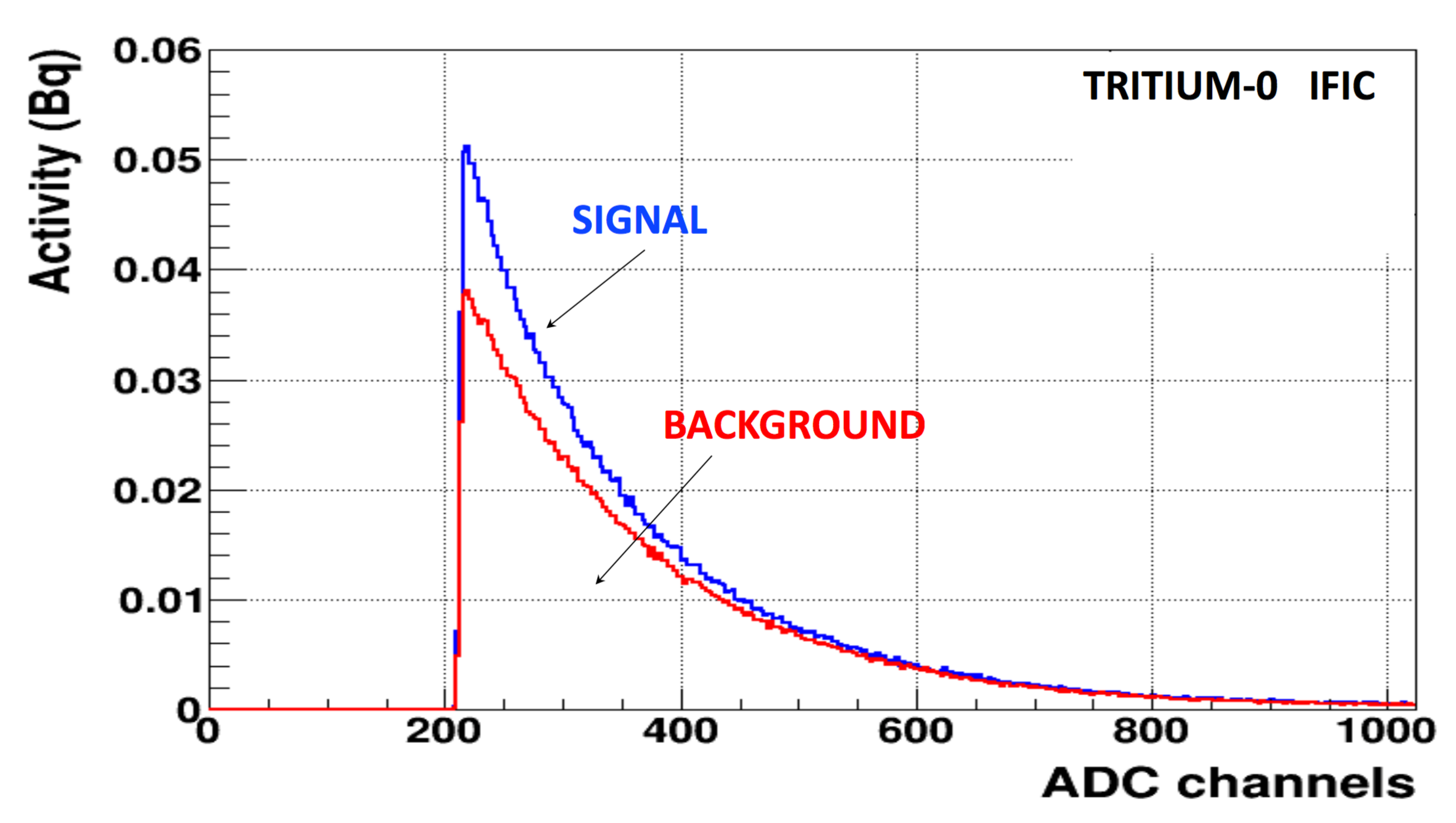}
\caption{Data spectra of hyperpure water (red) and of hyperpure tritiated water (108.11 MBq/L, blue) in number of ADC channels, from TRITIUM-0 prototype.}
\label{figure-9}
\end{figure}

A clear signal of tritium was observed in the proof-of-concept prototype TRITIUM-0 as shown in Fig.~\ref{figure-9}, in which the signal from tritiated hyperpure water is observed (blue spectrum) above the background signal (red spectrum) from a similar volume of hyperpure water without tritium. However, the signal observed from highly tritiated water (108.11 MBq/L) was weak, clearly indicating poor light production, collection or transmission to the PMTs.  Indeed, among the issues identified to determine the tritium signal magnitude are the geometrical configuration of the fiber-bundle, the condition of the fibers surface, the water circulation pattern in the detector vessel, the reflectivity of the inner surfaces of the water-vessel, and most specially, the magnitude of the background signal in the detector.  

\begin{figure}[!t]
\centering
\includegraphics[width=3.5in]{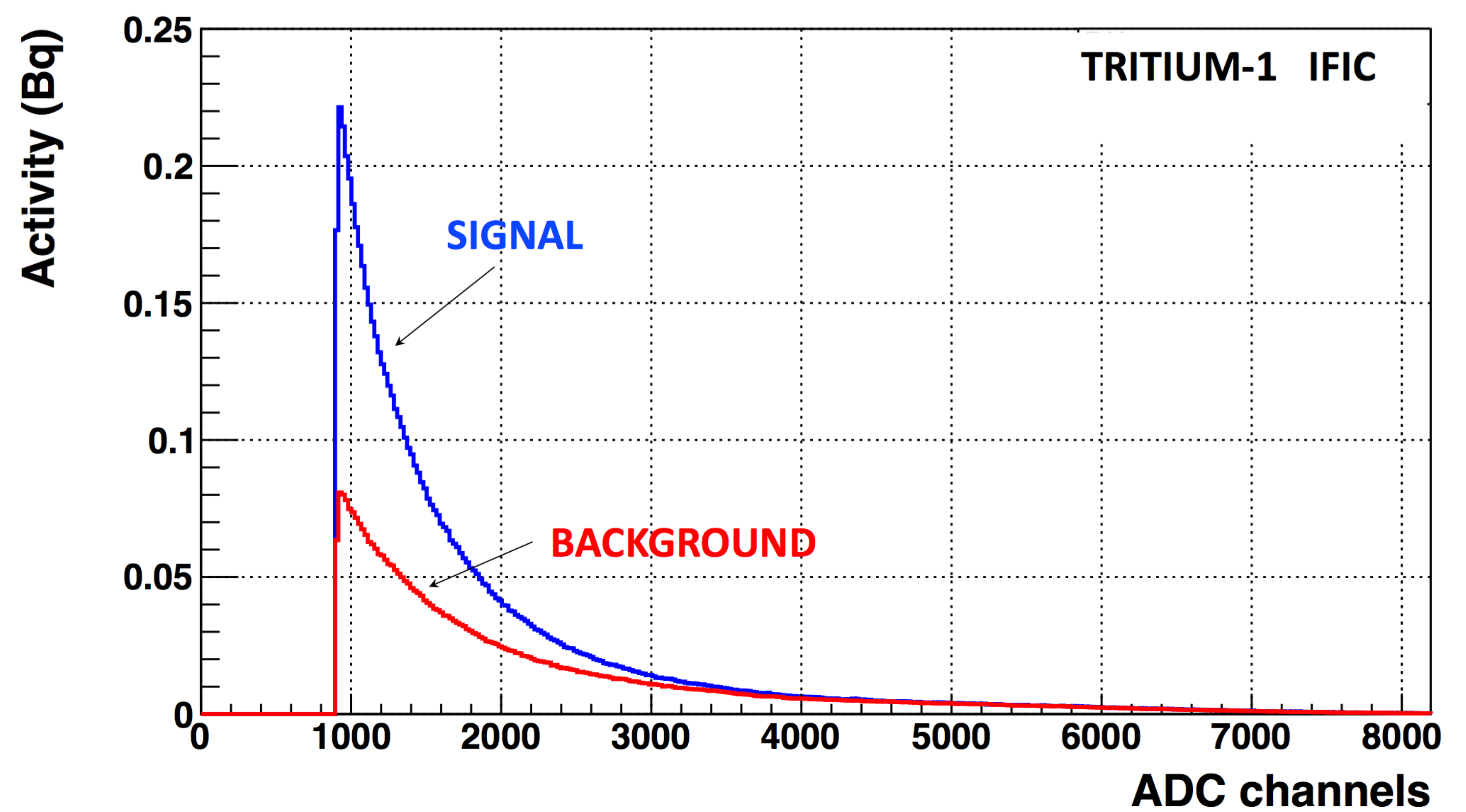}
\caption{Data spectra of hyperpure water (red) and of hyperpure tritiated water (108.11 MBq/L, blue) in number of ADC channels, from TRITIUM-1 IFIC prototype.}
\label{figure-10}
\end{figure}

In TRITIUM-1 prototypes the fibers have undergone a thorough conditioning protocol and were set up straight, avoiding light losses due to bendings and poor contact with water.  In Fig.~\ref{figure-10}, the signal from tritiated water with similar activity as for TRITIUM-0, was shown to improve the tritium signal by one order of magnitude compared to that measured in the previous prototype. TRITIUM-1 IFIC was also used to assess the stability of the fibers in water response over time. Through regular measurements in the laboratory, the detector prototype was proved to maintain a stable response over a period of 9 months. Further measurements are foreseen under shieldings in Almaraz TRITIUM site, to assess its increase of sensitivity with background rejection.     

\begin{figure}[!t]
\centering
\includegraphics[width=3.4in]{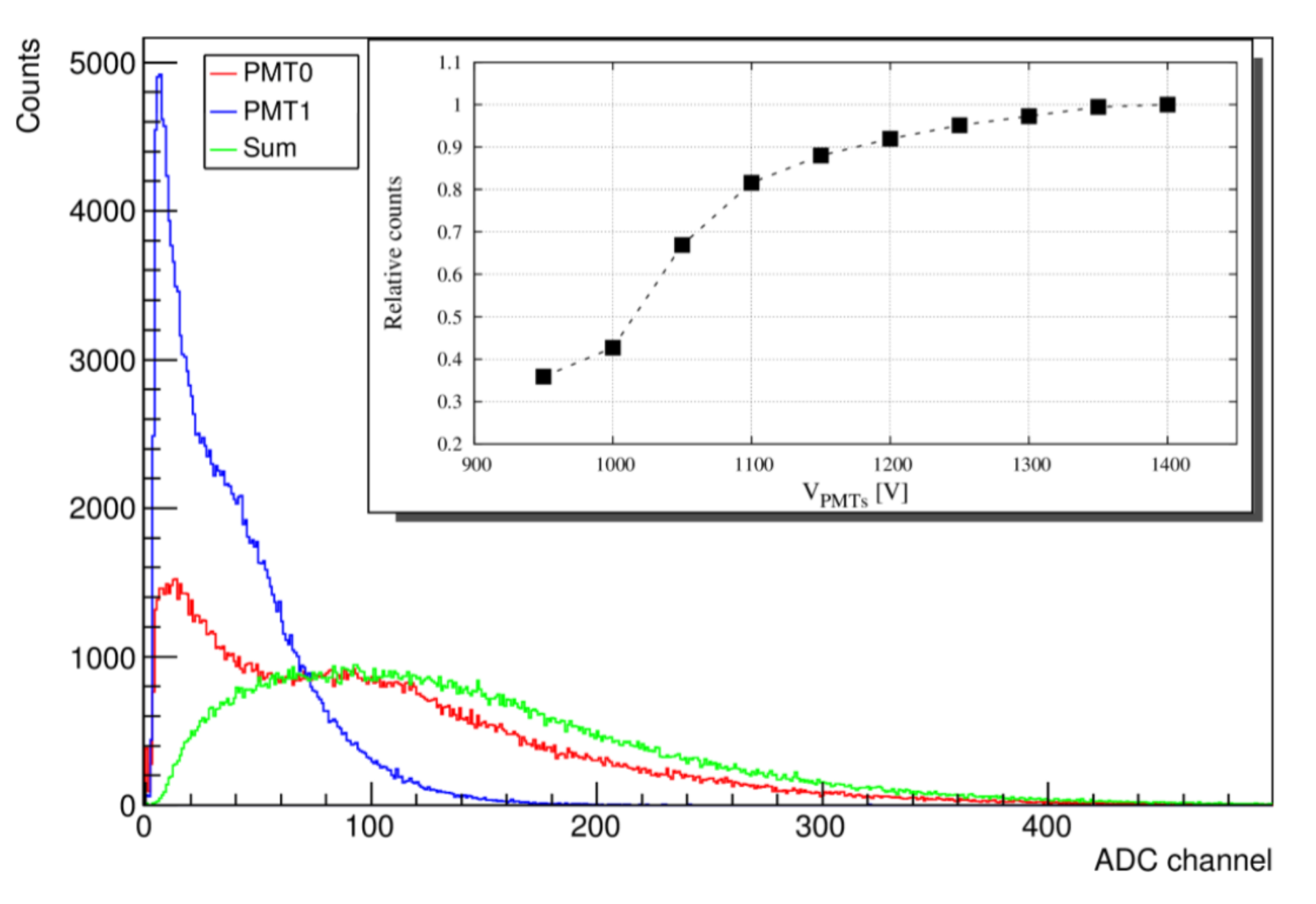}
\caption{Pulse-height distribution for each of the PMTs (V=-1250 V) of TRITIUM-1 Aveiro, and their sum. In the insert is the normalized count rate as a function of PMTs voltage when operating in coincidence mode. The dashed line is just a guide-to-the-eyes.}
\label{figure-11}
\end{figure}

In TRITIUM-1 Aveiro prototype, a substantial increase in tritium level sensitivity compared to the other prototypes is expected, due to the much higher number of fibers, thus much larger total detection surface, and also, due to the reduction of the PMTs noise with the implementation of a coincidence readout of the fibers. 
Before the use of tritiated water in this prototype, the capability of low energy detection by the scintillating fibers was first tested. For that, a small bundle of 12 Saint-Gobain BCF-12 fibers without cladding were assembled and wrapped in teflon tape. The fibers were positioned between 2 Hamamatsu R2154-02 PMTs acquiring in a coincidence window of 500 ns through a CAEN V1724 digitizer. The PMTs signals where pre-ampliﬁed by a Cremat Cr-110 module and the fibers where irradiated with 5.9 keV photons provided by a $^{55}$Fe source. A study of the number of detected events as a function of the PMTs voltage was carried out and the results are presented in the insert of Fig.~\ref{figure-11}. The increase of the number of events observed with the voltage increase in the PMTs is due to the increase of the PMTs detection efficiency, until a shape evidencing a plateau is reached around 1300 V. In background we can observe the pulse-height amplitude for both PMTs and the corresponding distribution of their sum. The different distributions for the PMTs are due to the difference in the intrinsic gain although they were fed with the same voltage (-1250V).
The in-water measurements with hyperpure water and afterwards hyperpure tritiated water in TRITIUM-1 Aveiro are currently in progress and will be reported in a further publication.

\section{Conclusion and Outlook}

The TRITIUM project aims at the design,  construction and commissioning of a near real-time automatic detector system for monitoring low levels of tritium in water in the vicinity of nuclear power plants. The tritium monitor has to measure tritium levels $< 100$ Bq/L, being this the limit required by the European Council Directive 2013/51/EURATOM for water for human consumption. The TRITIUM detector base design uses scintillating optical fibers read out by silicon photomultiplier arrays. 

GEANT4 simulations were carried out to optimize the detector design (scintillating fiber length, diameter, geometrical configuration) and several prototypes were developed and operated in the laboratories, with the aim of providing a proof of concept of the detector, and for studying its sensitivity and mechanical configuration. 
The simulations indicate that the deposited energy in fibers of 2 mm diameter without cladding will be peaked around 12 keV, with a distribution ranging from 2 to 18 keV. Moreover the electrons that reach the fiber surface will be created at a maximum distance of 5~$\mu$m from the fiber surface.
The possibility to use scintillating fibers to detect, in coincidence mode, low energies such as the 5.9 keV photons from a $^{55}$Fe source, was demonstrated as well as the scaling of the detector sensitivity with the increase of the number of scintillating fibers and the optimization of their surface conditions and geometrical configuration.

The results of this R\&D studies allowed the design of a modular TRITIUM detector, composed of 15 cylindrical cells containing each, 500 scintillating optical fibers of 25 cm long, 1 mm diameter, without cladding. Each cell will be instrumented with two SiPM arrays which read out the scintillation signals in coincidence mode to reject the intrinsic noise of the photodetectors. Additionally, active and passive shields based on scintillating plastics and lead bricks respectively, will be used for the rejection of cosmic and natural radioactivity backgrounds in the tritium detector. These shields are essential to reach high sensitivity to tritium levels below 100 Bq/L. Furthermore, a water purification system was developed to supply the detector with hyperpure water with conductivity close to 10~$\mu$S/cm. This system which has been shown to preserve the tritium level prior to the purification process, will provide a water flow that enables renewal of the detector water-content each 10 min for tritium measurement.

The construction of the full TRITIUM system is currently in progress. Its commissioning is foreseen in Spring 2019 in Arrocampo dam at Almaraz (Spain) nuclear power plant. 

\section*{Acknowledgment}

This work was supported by the INTERREG-SUDOE  EEC program through the project TRITIUM - SOE1/P4/E0214

\end{document}